\documentclass[a4paper, oneside, twocolumn, notitlepage, 10pt]{extarticle_ecoc}

\usepackage{ecoc}
\usepackage{sfmath}
\usepackage[per-mode=symbol,detect-all]{siunitx}
\usepackage[capitalise]{cleveref}
\usepackage[acronym,nomain]{glossaries}
\newcommand{\SetCapsType}{normalcaps}
\usepackage[UKenglish]{babel}
\usepackage{xspace}  
\usepackage[shortcuts]{extdash}
\usepackage{listofitems,pgffor}    
\usepackage{siunitx}
\usepackage{xstring}    
\usepackage{silence}
\usepackage{xparse}

\setsepchar{;}  

\ifdefined\silencecommonwarnings
\else
	\def\silencecommonwarnings{true} 
\fi

\ifbool{\silencecommonwarnings}{%
    \WarningFilter{ECOtools}{Cannot define: DH}%
    \WarningFilter{ECOtools}{Cannot define: PAM}%
    \WarningFilter{ECOtools}{Cannot define: QAM}%
    \WarningFilter{ECOtools}{Cannot define: SI}%
    \WarningFilter{ECOtools}{Cannot define: PV}%
    \WarningFilter{ECOtools}{Cannot define: LP}%
    \WarningFilter{ECOtools}{Cannot define: RN}%
    \WarningFilter{ECOtools}{Cannot define: uLP}%
    \WarningFilter{ECOtools}{Redefining DH}%
    }{}

\makeatletter
\providecommand{\SetCapsType}{smallcaps}

\long\def\@scTrue{smallcaps}
\long\def\@scFalse{normalcaps}
\newcommand{\acroSCaps}[1]{%
    \ifx\SetCapsType\@scTrue 
        \textsc{#1}%
    \else
        \MakeUppercase{#1}%
    \fi
}
\makeatother

\usepackage{scalerel}
\makeatletter
\newcommand\scslash{%
\ifx\SetCapsType\@scTrue 
    \protect\stretchrel*{$/$}{\textsc{e}}
\else
    /
\fi
} 
\makeatother 

\makeatletter
\@ifpackageloaded{babel}{%
    \newcommand{\usuk}[2]{%
        \iflanguage{USenglish}{#1}{#2}%
    }%
}{%
    \newcommand{\usuk}[2]{%
        #1%
    }%
}%

\newcommand{\langcheck}[2]{
    \@ifpackageloaded{babel}{%
        \iflanguage{USenglish}{#1}{#2}%
    }{%
        #1%
    }%
}

\makeatother

\newcommand{\short}[1]{%
    \glsentrytext{#1}\xspace%
}
\newcommand{\shortfakeplural}[1]{%
    \glsentrytext{#1}s\xspace%
}
\newcommand{\Short}[1]{%
    \Glsentrytext{#1}\xspace%
}
\newcommand{\normal}[1]{%
    \gls{#1}\xspace%
}
\newcommand{\longacr}[1]{%
    \acrlong{#1}\xspace%
}
\newcommand{\plural}[1]{%
    \glspl{#1}\xspace%
}
\newcommand{\full}[1]{%
    \acrfull{#1}\xspace%
}
\newcommand{\fullplural}[1]{%
    \acrfullpl{#1}\xspace%
}
\newcommand{\Normal}[1]{%
    \Gls{#1}\xspace%
}
\newcommand{\Plural}[1]{%
    \Glspl{#1}\xspace%
}
\newcommand{\Full}[1]{%
    \Acrfull{#1}\xspace%
}
\newcommand{\Fullplural}[1]{%
    \Acrfullpl{#1}\xspace%
} 

\newcommand{\texpdfif}[2]{%
    \ifcsname texorpdfstring\endcsname%
        \texorpdfstring{#1{#2}}{#2\xspace}%
    \else%
        #1{#2}%
    \fi%
}

\newcommand{\checkanddefine}[3]{%
	\ifcsname #1\endcsname%
        \PackageWarning{ECOtools}{Cannot define: #1 already defined, trying to define g#1 instead.}%
        \ifcsname g#1\endcsname%
            \PackageWarning{ECOtools}{Cannot define: g#1 also already defined.}%
    	\else%
        	\expandafter\newcommand\csname g#1\endcsname{%
        	    \texpdfif{#2}{#3}%
    	    }%
        \fi%
	\else%
    	\expandafter\newcommand\csname #1\endcsname{%
    	    \texpdfif{#2}{#3}%
	    }%
    \fi%
}

\newcommand{\redefine}[3]{%
    \PackageWarning{ECOtools}{Redefining #1}%
	\expandafter\renewcommand\csname #1\endcsname{%
	    \texpdfif{#2}{#3}%
    }%
}

\newcommand{\nAcronym}[4][]{%
	\newacronym[#1]{#2}{#3}{#4}%
	\checkanddefine{s#2}{\short}{#2}%
    \checkanddefine{s#2s}{\shortfakeplural}{#2}%
	\checkanddefine{#2}{\normal}{#2}%
	\checkanddefine{l#2}{\longacr}{#2}%
	\checkanddefine{#2s}{\plural}{#2}%
	\checkanddefine{f#2}{\full}{#2}%
	\checkanddefine{f#2s}{\fullplural}{#2}%
	\checkanddefine{su#2}{\Short}{#2}%
	\checkanddefine{u#2}{\Normal}{#2}%
	\checkanddefine{u#2s}{\Plural}{#2}%
	\checkanddefine{fu#2}{\Full}{#2}%
	\checkanddefine{fu#2s}{\Fullplural}{#2}%
	\IfStrEq{#2}{DH}{
	    \redefine{#2}{\normal}{#2}%
	    }{}%
}%

\NewDocumentCommand\qam{g}{%
    \IfNoValueTF{#1}{%
        \texpdfif{\gls}{QAM}\xspace%
        }{%
        \StrLen{#1}[\stringlength]%
        \ifnum\stringlength=0%
            \texpdfif{\gls}{QAM}\xspace%
        \else%
            {\qamlisthelper{#1}}%
        \fi%
        }%
}

\let\QAM\qam

\DeclareRobustCommand\qamlisthelper[1]{%
    \readlist*\args{#1}%
    \acroSCaps{\args[1]\=/}%
    \ifnum\argslen = 2%
        { and \acroSCaps{\args[2]}\=/}%
    \fi%
    \ifnum\argslen > 2%
        \foreach \n in {2,...,\argslen}{%
            \ifnum\n = \argslen%
                {, and }%
            \else 
                {, }%
            \fi%
            {\acroSCaps{\args[\n]}\=/}%
        }%
    \fi%
    \ifglsused{QAM}%
        {}%
        {ary }%
    \texpdfif{\gls}{QAM}%
}%

\NewDocumentCommand\pam{g}{%
    \IfNoValueTF{#1}{%
        \texpdfif{\gls}{PAM}\xspace%
        }{%
        \StrLen{#1}[\stringlength]%
        \ifnum\stringlength=0%
            \texpdfif{\gls}{PAM}\xspace%
        \else%
            {\pamlisthelper{#1}}%
        \fi%
        }%
}

\DeclareRobustCommand\pamlisthelper[1]{%
    \readlist*\args{#1}%
    \ifglsused{PAM}{%
        \texpdfif{\gls}{PAM}%
        \acroSCaps{\=/\args[1]}%
        \ifnum\argslen = 2%
            { and \=/\acroSCaps{\args[2]}}%
        \fi%
        \ifnum\argslen > 2%
            \foreach \n in {2,...,\argslen}{%
                \ifnum\n = \argslen%
                    {, and }%
                \else%
                    {, }%
                \fi%
                {\=/\acroSCaps{\args[\n]}}%
            }%
        \fi%
    }{%
        \acroSCaps{\args[1]\=/}%
        \ifnum\argslen = 2%
            { and \acroSCaps{\args[2]}\=/}%
        \fi%
        \ifnum\argslen > 2%
            \foreach \n in {2,...,\argslen}{%
                \ifnum\n = \argslen%
                    {, and }%
                \else%
                    {, }%
                \fi
                {\acroSCaps{\args[\n]}\=/}%
            }%
        \fi%
        {ary }%
        \texpdfif{\gls}{PAM}%
    }%
}%

\NewDocumentCommand\lp{g}{%
    \IfNoValueTF{#1}{%
        \texpdfif{\normal}{LP}%
        }{%
        \StrLen{#1}[\stringlength]%
        \ifnum\stringlength=0%
            \texpdfif{\normal}{LP}%
        \else%
            \ifglsused{LP}{}{\texpdfif{\normal}{LP}\xspace}%
            \lplisthelper[lp]{#1}%
        \fi%
        }%
}

\NewDocumentCommand\ulp{g}{%
    \IfNoValueTF{#1}{%
        \texpdfif{\Normal}{LP}\xspace%
        }{%
        \StrLen{#1}[\stringlength]%
        \ifnum\stringlength=0%
            \texpdfif{\Normal}{LP}\xspace%
        \else%
            \ifglsused{LP}{%
                \lplisthelper[Lp]{#1}%
            }{%
                \texpdfif{\Normal}{LP}\xspace\lplisthelper[lp]{#1}%
            }%
        \fi%
        }%
}
\DeclareRobustCommand\lplisthelper[2][lp]{%
    \readlist*\args{#2}%
    \foreach \n in {1,...,\argslen}{%
        \ifnum \n > 1%
            \ifnum \argslen > 2%
                {, }%
            \else%
                { }%
            \fi%
        \fi%
        \ifnum \n = \argslen%
            \ifnum \argslen > 1%
                {and }%
            \fi%
        \fi%
        \ifnum \n = 1%
            {\acroSCaps{#1}}
        \else%
            {\acroSCaps{\MakeLowercase{#1}}}%
        \fi%
        {\textsubscript{\StrSplit{\args[\n]}{2}{\csA}{\csB}\acroSCaps{\csA}\csB}}
    }%
}%

\nAcronym{128SPQAM}{\acroSCaps{128-sp-16-qam}}{128-ary set-partitioning \QAM{16}}

\nAcronym{2A8PSK}{\acroSCaps{2a8psk}}{2-ary amplitude 8-ary phaseshift keying}

\nAcronym{3CCMCF}{\acroSCaps{3cc-mcf}}{3-core coupled-core multi-core fiber}

\nAcronym{4D}{\acroSCaps{4d}}{four-dimensional}
\nAcronym{4D64PRS}{\acroSCaps{4d-64prs}}{\usuk{four-dimensional 64-ary polarization-ring-switching}{four-dimensional 64-ary polarisation-ring-switching}}
\nAcronym{4DOS128}{\acroSCaps{4d-os128}}{four-dimensional orthant-symmetric 128-ary modulation format}

\nAcronym{5B4D2A8PSK}{\acroSCaps{5b4d-2a8psk}}{5-bit four-dimensional two-amplitude 8-ary phase-shift keying}

\nAcronym{6B4D2A8PSK}{\acroSCaps{6b4d-2a8psk}}{6-bit four-dimensional two-amplitude 8-ary phase-shift keying}

\nAcronym{7B4D2A8PSK}{\acroSCaps{7b4d-2a8psk}}{7-bit four-dimensional two-amplitude 8-ary phase-shift keying}

\nAcronym{8D}{\acroSCaps{8d}}{eight-dimensional}\nAcronym{8D2048PRS}{\acroSCaps{8d-2048prs}}{eight-dimensional 2048-ary polarization-ring-switching}
\nAcronym{8D2048PRST1}{\acroSCaps{8d-2048prs-t1}}{eight-dimensional 2048-ary polarization-ring-switching type 1}
\nAcronym{8D2048PRST2}{\acroSCaps{8d-2048prs-t2}}{eight-dimensional 2048-ary polarization-ring-switching type 2}
\nAcronym{8DAPSK}{\acroSCaps{8d-apsk}}{eight-dimensional amplitude-phase-shift keying}

\nAcronym{ABC}{\acroSCaps{abc}}{automatic bias control}
\nAcronym{AC}{\acroSCaps{ac}}{alternating current}
\nAcronym{ADC}{\acroSCaps{adc}}{\usuk{analog-to-digital converter}{analogue-to-digital converter}}
\nAcronym{AGC}{\acroSCaps{agc}}{automatic gain control}
\nAcronym{AIR}{\acroSCaps{air}}{achievable information rate}
\nAcronym{AMZI}{\acroSCaps{amzi}}{asymmetric Mach–Zehnder interferometer}
\nAcronym{AO}{\acroSCaps{ao}}{adaptive optics}
\nAcronym{AOM}{\acroSCaps{aom}}{acousto-optic modulator}
\nAcronym{APD}{\acroSCaps{apd}}{avalanche photodiode}
\nAcronym{API}{\acroSCaps{api}}{application programming interface}
\nAcronym{AR}{\acroSCaps{ar}}{achievable rate}
\nAcronym{ARRWG}{\acroSCaps{a}rr\acroSCaps{wg}}{arrayed-waveguide grating}
\nAcronym{ASE}{\acroSCaps{ase}}{amplified spontaneous emission}
\nAcronym{ASK}{\acroSCaps{ask}}{amplitude-shift keying}
\nAcronym{ASIC}{\acroSCaps{asic}}{application-specific integrated circuit}
\nAcronym{ATS}{\acroSCaps{ats}}{alignment tracking sensor}
\nAcronym{AWG}{\acroSCaps{awg}}{arbitrary-waveform generator}
\nAcronym{AWGN}{\acroSCaps{awgn}}{additive white Gaussian noise}

\nAcronym{BBU}{\acroSCaps{bbu}}{baseband unit}
\nAcronym{BCH}{\acroSCaps{bch}}{Bose-Chaudhuri-Hocquenghem}
\nAcronym{BER}{\acroSCaps{ber}}{bit error rate}
\nAcronym{BERT}{\acroSCaps{bert}}{bit error rate tester}
\nAcronym{BICM}{\acroSCaps{bicm}}{bit-interleaved coded modulation}
\nAcronym{BMD}{\acroSCaps{bmd}}{bit-metric decoding}
\nAcronym{BPD}{\acroSCaps{bpd}}{balanced photo-diode}
\nAcronym{BPF}{\acroSCaps{bpf}}{bandpass filter}
\nAcronym{BPS}{\acroSCaps{bps}}{blind phase search}
\nAcronym{BPSK}{\acroSCaps{bpsk}}{binary phase-shift keying}
\nAcronym{BRGC}{\acroSCaps{brgc}}{binary reflected Gray code}
\nAcronym{BTB}{\acroSCaps{btb}}{back-to-back}

\nAcronym{CAGR}{\acroSCaps{cagr}}{compound annual growth rate}
\nAcronym{CCDM}{\acroSCaps{ccdm}}{constant composition distribution matching}
\langcheck{%
    \nAcronym{CCF}{\acroSCaps{ccf}}{coupled-core fiber}%
    }{%
    \nAcronym{CCF}{\acroSCaps{ccf}}{coupled-core fibre}%
}%
\nAcronym{CD}{\acroSCaps{cd}}{chromatic dispersion}
\nAcronym{CIR}{\acroSCaps{cir}}{channel impulse response}
\nAcronym{CMA}{\acroSCaps{cma}}{constant modulus algorithm}
\nAcronym{CMF}{\acroSCaps{cmf}}{core multiplicity factor}
\nAcronym{CMUX}{\acroSCaps{cmux}}{core multiplexer}
\nAcronym{COTS}{\acroSCaps{cots}}{commercial off-the-shelf}
\nAcronym{COW}{\acroSCaps{cow}}{coherent one-way}
\nAcronym{ChUT}{\acroSCaps{chut}}{channel under test}
\nAcronym[firstplural=channels under test (\acroSCaps{cut}s)]{CUT}{\acroSCaps{cut}}{channel under test}
\nAcronym{CRX}{\acroSCaps{crx}}{coherent receiver}
\nAcronym{CPE}{\acroSCaps{cpe}}{carrier phase estimation}
\nAcronym{CPU}{\acroSCaps{cpu}}{central processing unit}
\nAcronym{CSPR}{\acroSCaps{cspr}}{carrier-to-signal power ratio}
\nAcronym{CUDA}{\acroSCaps{cuda}}{compute unified device architecture}
\nAcronym{CVQKD}{\acroSCaps{cv-qkd}}{continuous-variable quantum key distribution}
\nAcronym{CW}{\acroSCaps{cw}}{continuous wave}
\nAcronym{CCD}{\acroSCaps{ccd}}{charge-coupled device}

\nAcronym{DA}{\acroSCaps{da}}{driver amplifier}
\nAcronym{DAC}{\acroSCaps{dac}}{\usuk{digital-to-analog converter}{digital-to-analogue converter}}
\nAcronym{DC}{\acroSCaps{dc}}{direct current}
\nAcronym{DBP}{\acroSCaps{dbp}}{digital backpropagation}
\nAcronym{DCF}{\acroSCaps{dcf}}{\usuk{dispersion-compensating fiber}{dispersion-compensating fibre}}
\langcheck{%
    \nAcronym{DCI}{\acroSCaps{dci}}{data center interconnect}
    }{%
    \nAcronym{DCI}{\acroSCaps{dci}}{data centre interconnect}
}%
\nAcronym{DDLMS}{\acroSCaps{dd-lms}}{decision-directed least mean square}
\nAcronym{DEMUX}{\acroSCaps{demux}}{de-multiplexer}
\nAcronym{DFA}{\acroSCaps{dfa}}{\usuk{doped fiber amplifier}{doped fibre amplifier}}
\nAcronym{DFB}{\acroSCaps{dfb}}{distributed feedback}
\nAcronym{DGD}{\acroSCaps{dgd}}{differential group delay}
\nAcronym{DH}{\acroSCaps{dh}}{digital holography}
\nAcronym{DM}{\acroSCaps{dm}}{distribution matcher}
\nAcronym{DMR}{\acroSCaps{dm}}{dichroic mirror}
\nAcronym{DMA}{\acroSCaps{dma}}{direct memory access}
\nAcronym{DMD}{\acroSCaps{dmd}}{differential mode delay}
\nAcronym{DMG}{\acroSCaps{dmg}}{differential modal gain}
\nAcronym{DMGD}{\acroSCaps{dmgd}}{differential mode group delay}
\nAcronym{DML}{\acroSCaps{dml}}{directly-modulated laser}
\nAcronym{DP}{\acroSCaps{dp}}{\usuk{dual-polarization}{dual-polarisation}}
\nAcronym{DPC}{\acroSCaps{dpc}}{digital pre-compensation}
\nAcronym{DPE}{\acroSCaps{dpe}}{digital pre-emphasis}
\nAcronym{DPIQ}{\acroSCaps{dp-iqm}}{\usuk{dual-polarization \acroSCaps{iq}-modulator}{dual-polarisation \acroSCaps{iq}-modulator}}
\nAcronym{DPLL}{\acroSCaps{dpll}}{digital phase-locked loop}
\nAcronym{DPS}{\acroSCaps{dps}}{differential phase shift}
\nAcronym{DQPSK}{\acroSCaps{dqpsk}}{differential quaternary phase-shift-keying}
\nAcronym{DRA}{\acroSCaps{dra}}{distributed Raman amplifier}
\nAcronym{DRE}{\acroSCaps{dre}}{digital resolution enhancer}
\nAcronym{DSB}{\acroSCaps{dsb}}{double-sideband}
\nAcronym{DSF}{\acroSCaps{dsf}}{\usuk{dispersion-shifted fiber}{dispersion-shifted fibre}} 
\nAcronym{DSO}{\acroSCaps{dso}}{digital sampling oscilloscope}
\nAcronym{DSP}{\acroSCaps{dsp}}{digital signal processing}
\nAcronym{DUT}{\acroSCaps{dut}}{device-under-test}

\nAcronym{DVQKD}{\acroSCaps{dv-qkd}}{discrete-variable quantum key distribution}

\nAcronym{DWDM}{\acroSCaps{dwdm}}{dense wavelength-division multiplexing}

\nAcronym{EAM}{\acroSCaps{eam}}{electro-absorption modulator}
\nAcronym{ECL}{\acroSCaps{ecl}}{external cavity laser}
\nAcronym{ED}{\acroSCaps{ed}}{Eucledian distance}
\nAcronym{EDF}{\acroSCaps{edf}}{\usuk{erbium-doped fiber}{erbium-doped fibre}}
\nAcronym{EDFA}{\acroSCaps{edfa}}{\usuk{erbium-doped fiber amplifier}{erbium-doped fibre amplifier}}
\nAcronym{ENOB}{\acroSCaps{enob}}{effective number of bits}
\nAcronym{ER}{\acroSCaps{er}}{extinction ratio}
\nAcronym{ESS}{\acroSCaps{ess}}{enumerative sphere shaping}

\langcheck{%
    \nAcronym{FBG}{\acroSCaps{fbg}}{fiber Bragg grating}%
    }{%
    \nAcronym{FBG}{\acroSCaps{fbg}}{fibre Bragg grating}%
}%
\nAcronym{FD}{\acroSCaps{fd}}{frequency domain}
\nAcronym{FDE}{\acroSCaps{fde}}{\usuk{frequency domain equalizer}{frequency domain equaliser}}
\nAcronym{FEC}{\acroSCaps{fec}}{forward error correction}
\nAcronym{FFE}{\acroSCaps{ffe}}{\usuk{feed-forward equalizer}{feed-forward equaliser}}
\nAcronym{FFT}{\acroSCaps{fft}}{fast Fourier transform}
\nAcronym{FIR}{\acroSCaps{fir}}{finite impulse response}
\nAcronym{FLOPS}{\acroSCaps{flops}}{floating point operations per second}
\nAcronym{FMEDF}{\acroSCaps{fm-edf}}{\usuk{few-mode erbium-doped fiber}{few-mode erbium-doped fibre}}
\nAcronym{FMEDFA}{\acroSCaps{fm-edfa}}{\usuk{few-mode erbium-doped fiber amplifier}{few-mode erbium-doped fibre amplifier}}
\langcheck{%
    \nAcronym{FMF}{\acroSCaps{fmf}}{few-mode fiber}%
    }{%
    \nAcronym{FMF}{\acroSCaps{fmf}}{few-mode fibre}%
}%
\nAcronym[plural=FM-MCF, firstplural=\usuk{few-mode multi-core fibers}{few-mode multi-core fibres}]{FMMCF}{\acroSCaps{fm-mcf}}{\usuk{few-mode multi-core fiber}{few-mode multi-core fibre}}
\langcheck{%
    \nAcronym{FMPBGF}{\acroSCaps{fm-pbgf}}{few-mode photonic bandgap fiber}%
    }{%
    \nAcronym{FMPBGF}{\acroSCaps{fm-pbgf}}{few-mode photonic bandgap fibre}%
}%
\nAcronym{FOV}{\acroSCaps{fov}}{field of view}
\nAcronym{FPGA}{\acroSCaps{fpga}}{field-programmable gate array}
\nAcronym{FWM}{\acroSCaps{fwm}}{four-wave mixing}
\nAcronym{FSO}{\acroSCaps{fso}}{free-space optical}
\nAcronym{FUT}{\acroSCaps{fut}}{\usuk{fiber under test}{fibre under test}}

\nAcronym{GD}{\acroSCaps{gd}}{group delay}
\nAcronym{GI}{\acroSCaps{gi}}{graded-index}
\nAcronym{GFF}{\acroSCaps{gff}}{gain flattening filter}
\nAcronym{GIFMF}{\acroSCaps{gi-fmf}}{\usuk{graded-index few-mode fiber}{graded-index few-mode fibre}}
\nAcronym{GIMMF}{\acroSCaps{gi-mmf}}{\usuk{graded-index multi-mode fiber}{graded-index multi-mode fibre}}
\nAcronym{GMI}{\acroSCaps{gmi}}{\usuk{generalized mutual information}{generalised mutual information}}
\nAcronym{GNSE}{\acroSCaps{gnse}}{generalized nonlinear Schr\"{o}dinger equation}
\nAcronym{GPU}{\acroSCaps{gpu}}{graphics processing unit}
\nAcronym{GS}{\acroSCaps{gs}}{geometric shaping}
\nAcronym{GV}{\acroSCaps{gv}}{group velocity}
\nAcronym{GVD}{\acroSCaps{gvd}}{group velocity dispersion}
\nAcronym{GPIO}{\acroSCaps{gpio}}{general purpose input output}
\nAcronym{GUI}{\acroSCaps{gui}}{graphical user interface}

\langcheck{%
    \nAcronym{HCF}{\acroSCaps{hcf}}{hollow-core fiber}
    }{%
    \nAcronym{HCF}{\acroSCaps{hcf}}{hollow-core fibre}
}%
\nAcronym{HDFEC}{\acroSCaps{hd-fec}}{hard-decision forward error correction}
\nAcronym{HG}{\acroSCaps{hg}}{Hermite-Gaussian}
\nAcronym{HOM}{\acroSCaps{hom}}{higher-order modes}
\nAcronym{HV}{\acroSCaps{hv}}{Hufnagel-Valley}
\nAcronym{HAP}{\acroSCaps{hap}}{Hufnagel-Andrew-Phillips}

\nAcronym{ICS}{\acroSCaps{ics}}{inter-core skew}
\nAcronym{ICXT}{\acroSCaps{ic-xt}}{inter-core cross-talk}
\nAcronym[plural=IL, firstplural=insertion losses (\acroSCaps{il})]{IL}{\acroSCaps{il}}{insertion loss}
\nAcronym{IFFT}{\acroSCaps{ifft}}{inverse fast Fourier transform}
\nAcronym{IIR}{\acroSCaps{iir}}{intensity impulse response}
\nAcronym{IM}{\acroSCaps{im}}{intensity modulator}
\nAcronym{IMDD}{\acroSCaps{im}\scslash \acroSCaps{dd}}{intensity-modulation direct-detection}
\nAcronym{IQM}{\acroSCaps{iqm}}{in-phase and quadrature modulator}
\nAcronym{ISI}{\acroSCaps{isi}}{inter-symbol interference}
\nAcronym{IP}{\acroSCaps{ip}}{intellectual property}

\nAcronym{JGN}{\acroSCaps{jgn}}{Japan Gigabit Network}

\nAcronym{KK}{\acroSCaps{kk}}{Kramers-Kronig}
\nAcronym{KIT}{\acroSCaps{kit}}{Karlsruhe Institute of Technology}

\nAcronym{LCOS}{\acroSCaps{LCoS}}{liquid crystal on silicon}
\nAcronym{LDPC}{\acroSCaps{ldpc}}{low-density parity-check}
\nAcronym{LEAF}{\acroSCaps{leaf}}{\usuk{large effective area fiber}{large effective area fibre}}
\nAcronym{LFSR}{\acroSCaps{lfsr}}{linear-feedback shift register}
\nAcronym{LG}{\acroSCaps{lg}}{Laguerre-Gaussian}
\nAcronym{LMS}{\acroSCaps{lms}}{least mean square}
\nAcronym{LLR}{\acroSCaps{llr}}{log-likelihood ratio}
\nAcronym{LO}{\acroSCaps{lo}}{local oscillator}
\nAcronym{LP}{\acroSCaps{lp}}{\usuk{linearly polarized}{linearly polarised}}
\nAcronym{LSPS}{\acroSCaps{lsps}}{\usuk{loop-synchronized polarization scrambler}{loop-synchronised polarisation scrambler}}
\nAcronym{LUT}{\acroSCaps{lut}}{lookup table}

\nAcronym{MVM}{\acroSCaps{mvm}}{matrix-vector multiplication}
\nAcronym{MB}{\acroSCaps{mb}}{Maxwell-Bolzmann}
\langcheck{%
    \nAcronym{MCF}{\acroSCaps{mcf}}{multi-core fiber}%
    }{%
    \nAcronym{MCF}{\acroSCaps{mcf}}{multi-core fibre}%
}%
\nAcronym{MDG}{\acroSCaps{mdg}}{mode dependent gain}
\nAcronym[firstplural=mode-dependent losses (\acroSCaps{mdl})]{MDL}{\acroSCaps{mdl}}{mode-dependent loss}
\nAcronym{MDM}{\acroSCaps{mdm}}{mode-division multiplexing}
\nAcronym{MEMS}{\acroSCaps{mems}}{micro-electro-mechanical systems}
\nAcronym{MF}{\acroSCaps{mf}}{matched filter}
\nAcronym{MFD}{\acroSCaps{mfd}}{mode field diameter}
\nAcronym{MI}{\acroSCaps{mi}}{mutual information}
\nAcronym{MIMO}{\acroSCaps{mimo}}{multiple-input multiple-output}
\nAcronym{ML}{\acroSCaps{ml}}{machine learning}
\nAcronym{MMA}{\acroSCaps{mma}}{multi-modulus algorithm}
\nAcronym{MMEDF}{\acroSCaps{mmedf}}{\usuk{multi-mode erbium-doped fiber}{multi-mode erbium-doped fibre}}
\nAcronym{MMEDFA}{\acroSCaps{mmedfa}}{\usuk{multi-mode erbium-doped fiber amplifier}{multi-mode erbium-doped fibre amplifier}}
\nAcronym{MMF}{\acroSCaps{mmf}}{\usuk{multi-mode fiber}{multi-mode fibre}}
\nAcronym{MMSE}{\acroSCaps{mmse}}{minimum mean squared error}
\nAcronym{MP}{\acroSCaps{mp}}{minimum phase}
\nAcronym{MPLC}{\acroSCaps{mplc}}{multi-plane light converter}
\nAcronym{MRC}{\acroSCaps{mrc}}{maximum ratio combining}
\nAcronym{MSE}{\acroSCaps{mse}}{mean squared error}
\nAcronym{MUX}{\acroSCaps{mux}}{multiplexer}
\nAcronym{MZM}{\acroSCaps{mzm}}{Mach-Zehnder modulator}
\nAcronym{MZI}{\acroSCaps{mzi}}{Mach-Zehnder interferometer}

\nAcronym{NA}{\acroSCaps{na}}{numerical aperture}
\langcheck{%
    \nAcronym{NANF}{\acroSCaps{nanf}}{nested antiresonant nodeless fiber}%
    }{%
    \nAcronym{NANF}{\acroSCaps{nanf}}{nested antiresonant nodeless fibre}%
}%
\nAcronym{NF}{\acroSCaps{nf}}{noise figure}
\nAcronym{NGMI}{\acroSCaps{ngmi}}{\usuk{normalized generalized mutual information}{normalised generalised mutual information}}
\nAcronym{NLSE}{\acroSCaps{nlse}}{nonlinear Schr\"{o}ding equation}
\nAcronym{NN}{\acroSCaps{nn}}{neural network}
\nAcronym{NIC}{\acroSCaps{nic}}{network interface card}
\nAcronym{NICT}{\acroSCaps{nict}}{National Institute of Information and Communications Technology}
\nAcronym{NIR}{\acroSCaps{nir}}{near-infrared}
\nAcronym{NISTSTS}{\acroSCaps{nist-sts}}{National Insitute of Standards and Technology: Statistical Test Suite}
\nAcronym{NRZ}{\acroSCaps{nrz}}{non-return-to-zero}
\nAcronym{NZDSF}{\acroSCaps{nz-dsf}}{\usuk{non-zero dispersion-shifted fiber}{non-zero dispersion-shifted fibre}} 

\nAcronym{OAM}{\acroSCaps{oam}}{orbital angular momentum}
\nAcronym{OBTB}{\acroSCaps{obtb}}{optical back-to-back}
\nAcronym{OCT}{\acroSCaps{oct}}{outer cladding thickness}
\nAcronym{ODE}{\acroSCaps{ode}}{ordinary differential equation}
\nAcronym{ODL}{\acroSCaps{odl}}{optical delay line}
\nAcronym{OEO}{\acroSCaps{oeo}}{optical-electrical-optical}
\nAcronym{OFC}{\acroSCaps{ofc}}{Optical Fiber Communications Conference}
\nAcronym{OFDR}{\acroSCaps{ofdr}}{optical frequency-domain reflectometer} 
\nAcronym{OFDM}{\acroSCaps{ofdm}}{orthogonal frequency division multiplexing}
\nAcronym{OH}{\acroSCaps{oh}}{overhead}
\nAcronym{OMFT}{\acroSCaps{omft}}{optical multi-format transmitter}
\nAcronym{OOK}{\acroSCaps{ook}}{on-off keying}
\nAcronym{OP}{\acroSCaps{op}}{optical processor}
\nAcronym{OPLL}{\acroSCaps{opll}}{optical phase-locked loop}
\nAcronym{OSA}{\acroSCaps{osa}}{\usuk{optical spectrum analyzer}{optical spectrum analyser}}
\nAcronym{OSNR}{\acroSCaps{osnr}}{optical signal-to-noise ratio}
\nAcronym{OTDR}{\acroSCaps{otdr}}{optical time-domain reflectometer}
\nAcronym{OTF}{\acroSCaps{otf}}{optical tunable filter}
\langcheck{%
    \nAcronym{OVNA}{\acroSCaps{ovna}}{optical vector network analyzer}%
    }{%
    \nAcronym{OVNA}{\acroSCaps{ovna}}{optical vector network analyser}%
}%
\nAcronym{OTG}{\acroSCaps{otg}}{optical turbulence generator}

\nAcronym{PAM}{\acroSCaps{pam}}{pulse-amplitude modulation}
\nAcronym{PAS}{\acroSCaps{pas}}{probabilistic amplitude shaping}
\nAcronym{PAPR}{\acroSCaps{papr}}{peak-to-average power ratio}
\nAcronym{PBC}{\acroSCaps{pbc}}{\usuk{polarization beam combiner}{polarisation beam combiner}}
\langcheck{%
    \nAcronym{PBGF}{\acroSCaps{pbgf}}{photonic bandgap fiber}%
    }{%
    \nAcronym{PBGF}{\acroSCaps{pbgf}}{photonic bandgap fibre}%
}%
\nAcronym{PBS}{\acroSCaps{pbs}}{polarization beam splitter}
\nAcronym{PC}{\acroSCaps{pc}}{physical contact}
\nAcronym{PCVD}{\acroSCaps{pcvd}}{plasma chemical vapor depostion}
\nAcronym{PCG}{\acroSCaps{pcg64}}{64-bit permuted congruential generator}
\nAcronym{PD}{\acroSCaps{pd}}{photodiode}
\nAcronym{PDF}{\acroSCaps{pdf}}{probability density function}
\langcheck{%
    \nAcronym{PDL}{\acroSCaps{pdl}}{polarization-dependent loss}
    }{%
    \nAcronym{PDL}{\acroSCaps{pdl}}{polarisation-dependent loss}
}%
\nAcronym{PDM}{\acroSCaps{pdm}}{\usuk{polarization-division multiplexing}{polarisation-division multiplexing}}
\langcheck{%
    \nAcronym{PER}{\acroSCaps{per}}{polarization extinction ratio}%
    }{%
    \nAcronym{PER}{\acroSCaps{per}}{polarisation extinction ratio}%
}%
\nAcronym{PIC}{\acroSCaps{pic}}{photonic integrated circuit}
\nAcronym{PL}{\acroSCaps{pl}}{photonic lantern}
\nAcronym{PM}{\acroSCaps{pm}}{polarization-multiplexed}
\nAcronym{PMBPSK}{\acroSCaps{pm-bpsk}}{polarization-multiplexed binary phase-shift keying}
\nAcronym{PMQPSK}{\acroSCaps{pm-qpsk}}{polarization-multiplexed quaternary phase-shift keying}
\nAcronym{PM8QAM}{\acroSCaps{pm-8qam}}{polarization-multiplexed 8-ary quadrature amplitude modulation}
\nAcronym{PMD}{\acroSCaps{pmd}}{\usuk{polarization mode dispersion}{polarisation mode dispersion}}
\langcheck{%
    \nAcronym{PMF}{\acroSCaps{pmf}}{polarization-maintaining fiber}%
    }{%
    \nAcronym{PMF}{\acroSCaps{pmf}}{polarization-maintaining fibre}%
}%
\nAcronym{PMP}{\acroSCaps{pmp}}{phase-matching point}
\nAcronym{PNOB}{\acroSCaps{pnob}}{physical number of bits}
\nAcronym{PON}{\acroSCaps{pon}}{passive-optical network}
\nAcronym{PRBS}{\acroSCaps{prbs}}{pseudorandom bit sequence}
\nAcronym{PRNG}{\acroSCaps{PRNG}}{pseudo random number generator}
\nAcronym{PROFA}{\acroSCaps{profa}}{\usuk{pitch reducing optical fiber array}{pitch reducing optical fibre array}}
\nAcronym{PPM}{\acroSCaps{ppm}}{pulse-position modulation}
\nAcronym{PS}{\acroSCaps{ps}}{probabilistic shaping}
\langcheck{%
    \nAcronym{PSCF}{\acroSCaps{pscf}}{pure silica core fiber}%
    }{%
    \nAcronym{PSCF}{\acroSCaps{pscf}}{pure silica core fibre}%
}%
\nAcronym{PSD}{\acroSCaps{psd}}{power spectral density}
\nAcronym{PSF}{\acroSCaps{psf}}{point spread function}
\nAcronym{PSK}{\acroSCaps{psk}}{phase-shift keying}
\nAcronym{PSP}{\acroSCaps{psp}}{\usuk{principal states of polarization}{principal states of polarisation}}
\nAcronym{PSW}{\acroSCaps{psw}}{\usuk{polarization switch}{polarisation switch}}
\nAcronym{PID}{\acroSCaps{pid}}{proportional–integral–derivative}
\nAcronym{PV}{\acroSCaps{pv}}{process value}

\nAcronym{QAM}{\acroSCaps{qam}}{quadrature amplitude modulation}
\nAcronym{QBER}{\acroSCaps{qber}}{quantum bit error rate}
\nAcronym{QKD}{\acroSCaps{qkd}}{quantum key distribution}
\nAcronym{QPSK}{\acroSCaps{qpsk}}{quadrature phase-shift keying}
\nAcronym{QRNG}{\acroSCaps{qrng}}{quantum random number generator}
\nAcronym{QSM}{\acroSCaps{qsm}}{quasi-single-mode}

\nAcronym{RAM}{\acroSCaps{ram}}{random-access memory}
\nAcronym{RC}{\acroSCaps{rc}}{raised cosine}
\nAcronym{RCMF}{\acroSCaps{rcmf}}{relative core multiplicity factor}
\nAcronym{RCMCF}{\acroSCaps{rc-mcf}}{\usuk{randomly-coupled multi-core fiber}{randomly-coupled multi-core fibre}}
\nAcronym{RF}{\acroSCaps{rf}}{radio frequency}
\nAcronym{RFSoC}{\acroSCaps{rfsoc}}{radio frequency system-on-chip}
\nAcronym{RI}{\acroSCaps{ri}}{refractive index}
\nAcronym{RLS}{\acroSCaps{rls}}{recursive least squares}
\nAcronym{RRC}{\acroSCaps{rrc}}{root-raised-cosine}
\nAcronym{ROADM}{\acroSCaps{roadm}}{reconfigurable optical add-drop multiplexer}
\nAcronym{ROI}{\acroSCaps{roi}}{region of interest} 
\nAcronym{RTO}{\acroSCaps{rto}}{real-time oscilloscope} 
\nAcronym{RZDBPSK}{\acroSCaps{rz-dbpsk}}{return-to-zero differential binary phase-shift keying} 
\nAcronym{RZDQPSK}{\acroSCaps{rz-dqpsk}}{return-to-zero differential quaternary phase-shift keying}
\nAcronym{RN}{\acroSCaps{RN}}{random number}

\nAcronym{S2}{\acroSCaps{S\textsuperscript{2}}}{spatially and spectrally resolved}
\nAcronym{SA}{\acroSCaps{sa}}{simulated annealing}
\nAcronym{SamPerSym}{\acroSCaps{sps}}{samples per symbol}
\nAcronym{SBS}{\acroSCaps{sbs}}{stimulated Brillouin scattering}
\nAcronym{SCM}{\acroSCaps{scm}}{subcarrier multiplexing}
\nAcronym{SDFEC}{\acroSCaps{sd-fec}}{soft-decision forward error correction}
\nAcronym{SDM}{\acroSCaps{sdm}}{space-division multiplexing}
\nAcronym{SE}{\acroSCaps{se}}{spectral efficiency}
\nAcronym{SER}{\acroSCaps{ser}}{symbol error rate}
\nAcronym{SI}{\acroSCaps{si}}{step index}
\nAcronym{SIFMF}{\acroSCaps{si-fmf}}{\usuk{step-index few-mode fiber}{step-index few-mode fibre}}
\nAcronym{SISMF}{\acroSCaps{si-smf}}{\usuk{step-index single-mode fiber}{step-index single-mode fibre}}
\nAcronym{SLM}{\acroSCaps{slm}}{spatial light modulator}
\nAcronym{SKR}{\acroSCaps{skr}}{secret key rate}
\nAcronym{SMD}{\acroSCaps{smd}}{spatial-mode dispersion}
\nAcronym{SSC}{\acroSCaps{ssc}}{spot-size converter}
\nAcronym{SMF}{\acroSCaps{smf}}{\usuk{single-mode fiber}{single-mode fibre}}
\nAcronym{SMU}{\acroSCaps{smu}}{source measure unit}
\nAcronym{SMUX}{\acroSCaps{smux}}{spatial multiplexer}
\nAcronym{SNR}{\acroSCaps{snr}}{signal-to-noise ratio}
\nAcronym{SNU}{\acroSCaps{snu}}{shot-noise unit}
\nAcronym{SOA}{\acroSCaps{soa}}{semiconductor optical amplifier}
\nAcronym[firstplural=\usuk{states of polarization (\acroSCaps{sop})}{states of polarisation (\acroSCaps{sop})}]{SOP}{\acroSCaps{sop}}{\usuk{state of polarization}{state of polarisation}}
\nAcronym{SPM}{\acroSCaps{spm}}{self-phase modulation}
\nAcronym{SPD}{\acroSCaps{spd}}{single-photon detector}
\nAcronym{SPS}{\acroSCaps{sps}}{samples per symbol}
\nAcronym{SRS}{\acroSCaps{srs}}{stimulated Raman scattering}
\nAcronym{SSB}{\acroSCaps{ssb}}{single-sideband}
\nAcronym{SSBI}{\acroSCaps{ssbi}}{signal-signal beat interference}
\nAcronym{SSFM}{\acroSCaps{ssfm}}{split-step Fourier method}
\nAcronym{SSMF}{\acroSCaps{ssmf}}{\usuk{standard single-mode fiber}{standard single-mode fibre}}
\nAcronym{STAXT}{\acroSCaps{staxt}}{short-term average cross-talk}
\nAcronym{STL}{\acroSCaps{stl}}{swept tunable laser}
\nAcronym{SVD}{\acroSCaps{svd}}{singular value decomposition}
\nAcronym{SW}{\acroSCaps{sw}}{sequence-wise}
\nAcronym{SWI}{\acroSCaps{swi}}{swept wavelength interferometry}
\nAcronym{SP}{\acroSCaps{sp}}{setpoint}

\nAcronym{TAT}{\acroSCaps{tat}}{transatlantic}
\nAcronym{TC}{\acroSCaps{tc}}{tunable coupler}
\nAcronym{TD}{\acroSCaps{td}}{time domain}
\nAcronym{TDE}{\acroSCaps{tde}}{\usuk{time domain equalizer}{time domain equaliser}}
\nAcronym{TDFA}{\acroSCaps{tdfa}}{thulium doped-fiber amplifier}
\nAcronym{TDM}{\acroSCaps{tdm}}{time-domain multiplexing}
\nAcronym{TDMSDM}{\acroSCaps{tdm-sdm}}{time-domain multiplexed space-division multiplexing}
\nAcronym{TE}{\acroSCaps{te}}{transverse electric}
\nAcronym{TRNG}{\acroSCaps{TRNG}}{true random number generator}
\nAcronym{TEC}{\acroSCaps{tec}}{thermally-expanded-core}
\nAcronym{TOPS}{\acroSCaps{tops}}{thermo-optic phase shifter}
\nAcronym{TLS}{\acroSCaps{tls}}{tunable laser source}
\nAcronym{TIA}{\acroSCaps{tia}}{trans-impedance amplifier}
\nAcronym{TM}{\acroSCaps{tm}}{transverse magnetic}
\nAcronym{TH4D}{\acroSCaps{th-4d}}{time domain hybrid four-dimensional}
\nAcronym{TH4D2A8PSK}{\acroSCaps{th-4d-2a8psk}}{time-domain hybrid four-dimensional two-amplitude eight-phase-shift keying}
\nAcronym{TP}{\acroSCaps{tp}}{twisted pair}
\nAcronym{TTL}{\acroSCaps{ttl}}{transistor-transistor logic}

\nAcronym{UWB}{\acroSCaps{uwb}}{ultra-wideband}
\nAcronym{ULI}{\acroSCaps{uli}}{ultrafast laser inscription}

\nAcronym{VCSEL}{\acroSCaps{vcsel}}{vertical-cavity surface emitting laser}
\nAcronym{VHDL}{\acroSCaps{vhdl}}{\acroSCaps{Vhsic} Hardware Description Language}
\nAcronym{VOA}{\acroSCaps{voa}}{variable optical attenuator}

\nAcronym{WDL}{\acroSCaps{wdl}}{wavelength-dependent loss}
\nAcronym{WDM}{\acroSCaps{wdm}}{wavelength-division multiplexing}
\nAcronym{WFS}{\acroSCaps{wfs}}{wavefront sensor}
\nAcronym{WGA}{\acroSCaps{wga}}{weakly guiding approximation}
\nAcronym{WGN}{\acroSCaps{wgn}}{white Gaussian noise}
\nAcronym[longplural=wavelength selective switches]{WSS}{\acroSCaps{wss}}{wavelength selective switch}
\nAcronym{WC}{\acroSCaps{wc}}{weakly coupled}
\nAcronym{WCMCF}{\acroSCaps{wc-mcf}}{\usuk{weakly-coupled multi-core fiber}{weakly-coupled multi-core fibre}}

\nAcronym{XGM}{\acroSCaps{xgm}}{cross-gain modulation}
\nAcronym{XPM}{\acroSCaps{xpm}}{cross-phase modulation}
\nAcronym{XOR}{\acroSCaps{xor}}{exclusive or}
\langcheck{%
    \nAcronym{XPOLM}{\acroSCaps{xp}ol\acroSCaps{m}}{cross-polarization modulation}
    }{%
    \nAcronym{XPOLM}{\acroSCaps{xp}ol\acroSCaps{m}}{cross-polarisation modulation}
}%
\nAcronym{XT}{\acroSCaps{xt}}{cross-talk}

\nAcronym{3DWG}{\acroSCaps{3dwg}}{3D-waveguide}

\DeclareRobustCommand{\mkbibsuperscript}[1]{#1}
\DefineBibliographyExtras{english}{}

\begin{document}

\title{Compact Continuous-Variable Quantum Key Distribution System Employing Monolithically Integrated Silicon Photonic Transceiver}

\author{
   Denis Fatkhiev,
   João dos Reis Frazão,
   Alireza H. Derkani,
   Kadir G\" um\" u\c s,\\
   Menno van den Hout,
   Aaron Albores-Mejia,
   and Chigo Okonkwo
}

\maketitle

\begin{strip}
    \begin{author_descr}
    
        High Capacity Optical Transmission Lab, Eindhoven University of Technology, The Netherlands\\
        \textcolor{blue}{\uline{\href{mailto:d.fatkhiev@tue.nl}{d.fatkhiev@tue.nl}}}
    \end{author_descr}
\end{strip}

\renewcommand\footnotemark{}
\renewcommand\footnoterule{}

\begin{strip}
    \begin{ecoc_abstract}
        We demonstrate the first CV-QKD system featuring a custom-designed monolithic silicon photonic dual-polarisation transceiver. Leveraging PS-64-QAM, we achieved \SI{1.9}{Mbit/s} secret key rate across \SI{25}{km} of standard single-mode fibre, highlighting the potential of electronic-photonic integration for practical QKD. \textcopyright\,\,2025 The Authors
    \end{ecoc_abstract}
\end{strip}

\section{Introduction}

Owing to its high compatibility with conventional telecommunication hardware infrastructure, which enhances its practical appeal, and increasingly robust security proofs~\cite{leverrierComposableSecurityProof2015, pirandolaImprovedComposableKey2024}, \CVQKD is a promising technology for future information-theoretically secure communication networks.
While discrete-variable (DV) QKD uses single photons, \CVQKD offers advantages by employing coherent states and coherent detection, thus allowing the usage of components and \DSP techniques similar to coherent optical transmission systems.
Moreover, \CVQKD with discrete modulation significantly simplifies the modulation scheme, reducing the requirement for higher resolution \DACs~\cite{leverrierUnconditionalSecurityProof2009} and consequently lowering system complexity, thereby increasing the potential for its widespread adoption.

\begin{figure}[b!]
    \vspace{-5mm}
    \centering
    \includegraphics[width=72mm]{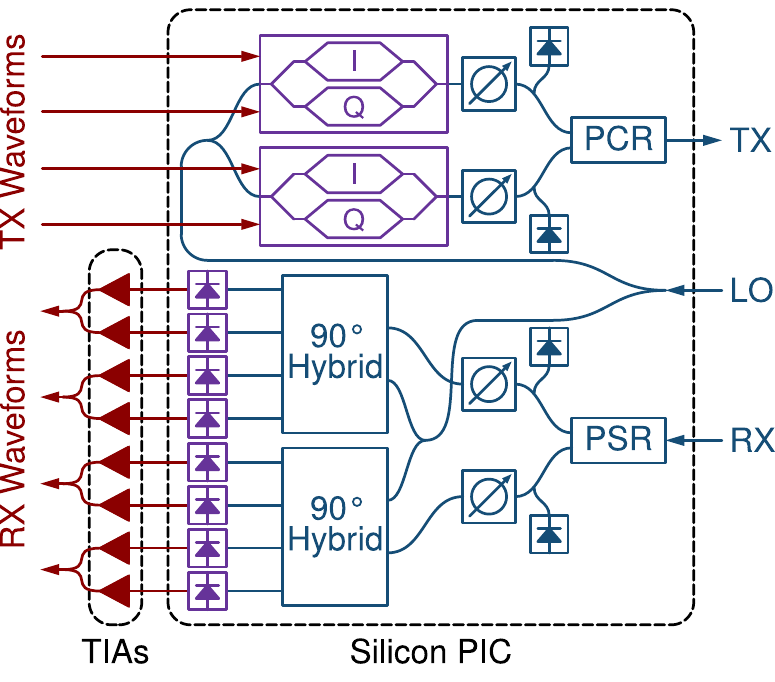}
    \caption{Schematic diagram of the transceiver chip.}
    \label{fig:trx}
\end{figure}

However, practical deployment has been hindered not only by the feasibility of specific components but also by the size, cost, and environmental sensitivity of bulk-optic implementations.
Therefore transition to \PICs is desirable for system miniaturisation, improved stability (especially critical for phase-sensitive \CVQKD systems), and achieving the economies of scale required for industrialisation~\cite{labonteIntegratedPhotonicsQuantum2024}. 
Recent developments in photonic integration for \CVQKD demonstrate the realisation of core system elements, including light sources \cite{liContinuousvariableQuantumKey2023}, single-polarisation~\cite{aldamaInPbasedCVQKDPIC2023, hajomer2025chipbased16gbaudcontinuousvariable} and dual-polarisation~\cite{ng2025gigabitratequantumkeydistribution} transmitters, and coherent receivers~\cite{pietriExperimentalDemonstrationContinuousvariable2024, ngChipintegratedHomodyneDetection2024}.
Although significant progress has been made, a complete, monolithic transceiver implementation leveraging polarisation multiplexing has not yet been demonstrated.

In this paper, we present the first \CVQKD system employing a custom-designed monolithically integrated silicon photonic dual-polarisation transceiver co-packaged with \TIAs.
Utilising polarisation and phase diversity coherent detection, probabilistically shaped (PS) 64-QAM modulation, and a tailored \DSP chain, the system reached \SI{1.9}{Mbit/s} \SKR across \SI{25}{km} of standard fibre.
For the security analysis, the asymptotic \SKR bound under arbitrary modulation~\cite{denysExplicitAsymptoticSecret2021} is applied, while the trusted noise assumption and finite-size effects are considered as described in~\cite{Jouguet}.
Our results demonstrate the potential roadmap towards practical and compact QKD systems exploiting advanced photonic-electronic integration.

\section{Photonic Front-End}

The functional schematic of the transceiver chip, detailing its core building blocks for both transmit and receive paths, is shown in \cref{fig:trx}.
The transmitter takes four electrical signals representing each polarisation state's in-phase (I) and quadrature (Q) components, which drive two independent IQ modulators.
High-speed modulation is based on carrier-depletion phase shifters within the Mach-Zehnder interferometers, supported by thermal phase shifters for bias control. 
An external \LO source provides the optical carrier, which is split into both modulators.
Note that the modulation stage is designed and optimised for driverless operation to reduce the system's excess noise.
Following modulation, two streams are combined with an integrated polarisation combiner-rotator (PCR).
Monitor \PDs provide feedback signals for \ABC to set and maintain the desired operating points of the modulators.
The receiver divides the signal into its two constituent orthogonal polarisation components with a polarisation splitter-rotator (PSR).
Coherent detection is performed for each polarisation with two integrated \ang{90}~hybrid mixers, each combining one polarisation component with a portion of the LO signal.
The outputs from the optical hybrids, corresponding to the I and Q components for each polarisation, are detected by high-speed \PDs.
\TIAs convert the incoming photocurrents into amplified voltage signals representing the four baseband waveforms.
Optical-to-electrical conversion for balanced coherent detection and monitoring utilises Ge-on-Si \PDs optimised for telecom wavelengths.

\section{System Implementation}

\begin{figure*}[t!]
    \centering
    \includegraphics[width=\textwidth]{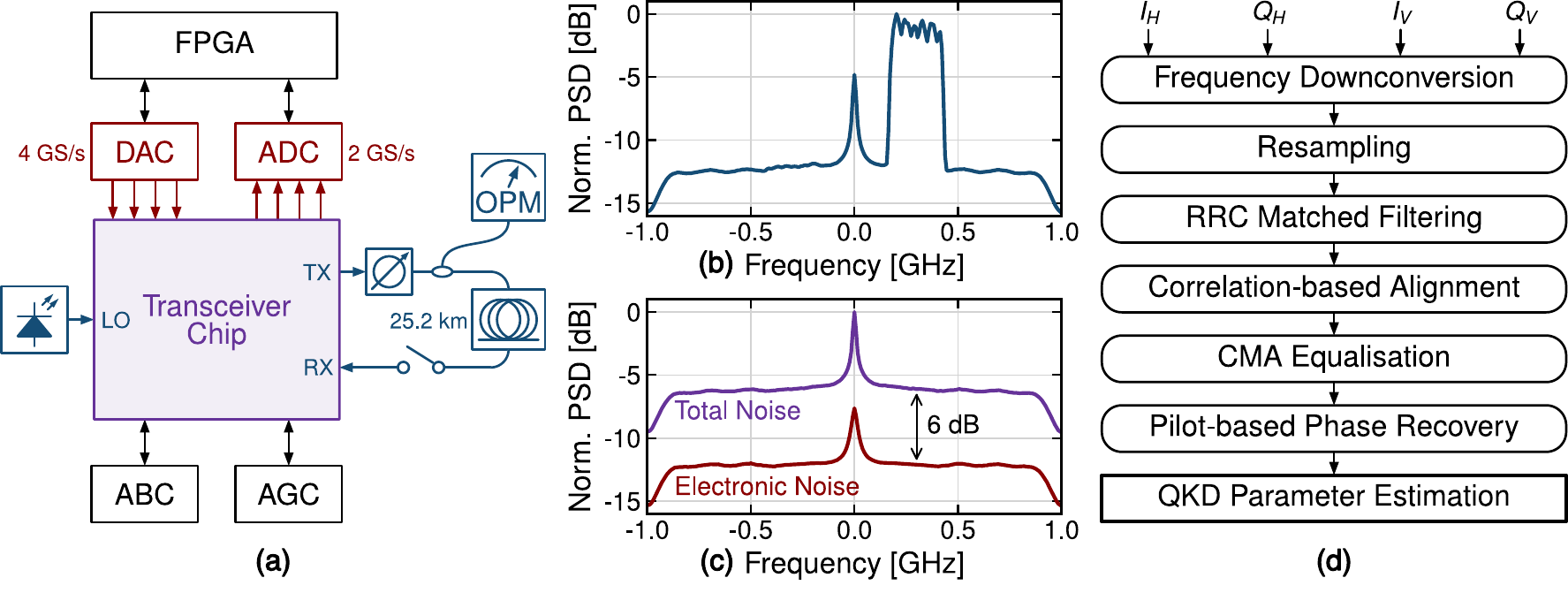}
    \caption{\textbf{(a)} Schematic of the experimental setup featuring transceiver and fibre link. \textbf{(b)} Normalised PSD estimate of the received signal. \textbf{(c)} Normalised PSD estimates comparing the electronic and total noise measured at the receiver. \textbf{(d)} Receiver DSP chain.}
    \label{fig:setup}
\end{figure*}

The experimental system, schematically detailed in \cref{fig:setup}a, centres around the silicon transceiver chip as the core photonic quantum engine; hence, the transmit and receive paths perform the roles of Alice and Bob, respectively.

\begin{figure}[b!]
    \vspace{-5mm}
    \centering
    \includegraphics[width=64mm]{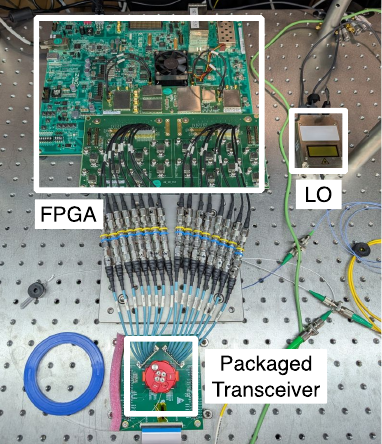}
    \caption{Photograph of the experimental setup, highlighting key hardware components.}
    \label{fig:photo}
\end{figure}

The signal generation process originates from the \FPGA, producing the digital baseband \SI{250}{MBaud} PS-64-QAM streams time-interleaved with QPSK pilot streams (power of pilot symbols \SI{15}{dB} higher than power of QKD symbols) in a 50:50 ratio, which are subsequently pulse-shaped using a \RRC filter with 10\% roll-off.
The \PSD estimate of the signal on the receiver side is presented on \cref{fig:setup}b.
To counteract signal quality deterioration induced by low-frequency noise at the receiver (see \cref{fig:setup}c), the signals are also up-converted by \SI{300}{MHz}.
These streams are fed to \SI{14}{bit} \DACs with a sampling rate of \SI{4}{GS/s}, to generate the analogue electrical waveforms for the modulators' inputs.
An external, narrow-linewidth laser operating at \SI{1550}{nm} provides the continuous-wave optical carrier.
The modulated optical signal exits the transmitter port and is attenuated by a \VOA for state preparation.
Part of the optical signal is tapped into an optical power meter (OPM) to estimate the modulation variance.
The signal is then transmitted over \SI{25.2}{km} of \SSMF, before being directed to the receiver.
An optical switch is placed before the receiver for the noise calibration~\cite{heideReceiverNoiseStability2023}.
After coherent detection, the analogue electrical signals are digitised by \SI{14}{bit} \ADCs with a sampling rate of \SI{2}{GS/s} and passed for further offline processing using the \DSP sequence outlined in \cref{fig:setup}d.
The hardware implementation is depicted in \cref{fig:photo}, demonstrating a practical and relatively compact experimental setup with its key components.
An additional control module, not shown in the figure, performs \ABC for the transmitter and \AGC for the receiver.

While based on conventional coherent communication techniques~\cite{farukDigitalSignalProcessing2017}, the implemented \DSP chain features some modifications.
Specifically, a short data-aided \LMS equalisation step precedes the \CMA equalisation stage to enhance convergence speed, and pilot symbols are used for carrier phase estimation.
Furthermore, the equaliser parameters (number of taps and step size) were optimised according to the approach proposed in~\cite{roumestanShapedConstellationContinuous2024}.
Due to the use of the same transceiver for transmit and receive in the experiment -- meaning that the carrier and \LO light stem from the same laser source -- the detection is effectively homodyne, and thus a distinct frequency recovery stage is omitted from the DSP chain.

\section{Results}

 \begin{figure*}[t!]
    \centering
    \includegraphics[width=\textwidth]{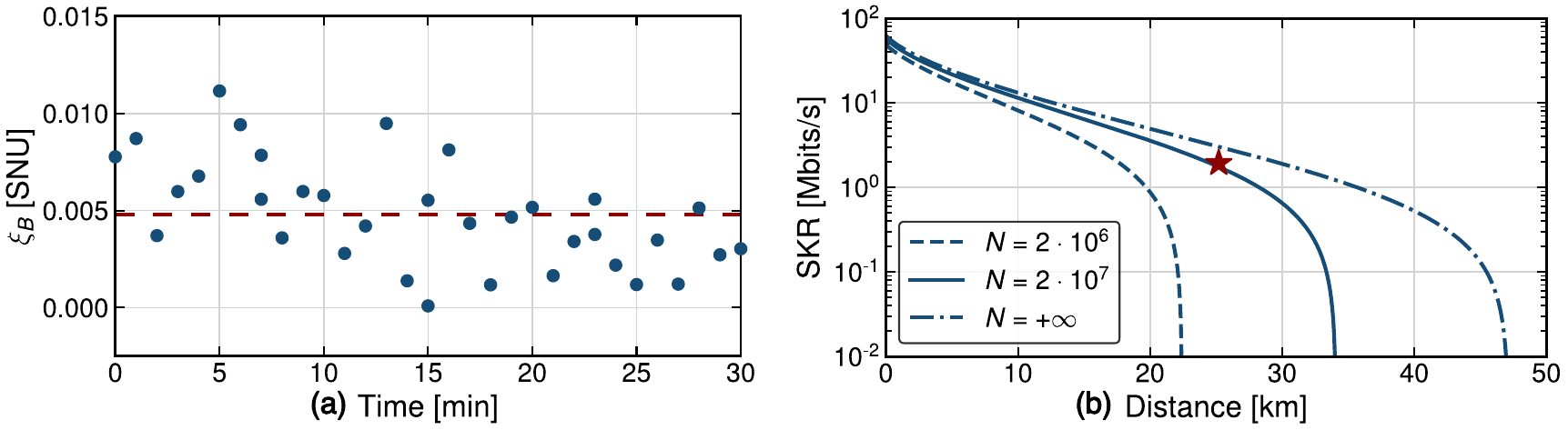}
    \caption{\textbf{(a)} Measured excess noise at the receiver over time, with the average value shown by the red dashed line.\\
    \textbf{(b)} \SKR versus transmission distance, illustrating theoretical performance for finite block sizes (N)\\
    and the asymptotic limit (N~$\rightarrow \infty$), highlighting the experimental result (red star).}
    \label{fig:results}
\end{figure*}

The system's performance was experimentally assessed through measurements of key parameters and estimations of the achievable \SKRs.
\Cref{fig:results}a presents the measured excess noise at Bob ($\xi_B$), expressed in shot noise units (SNU), monitored over 30 minutes.
The excess noise fluctuates around an average value of approximately \SI{0.005}{SNU}.
The Alice's modulation variance $V_A$ for the experiment was optimised and fixed to \SI{5.3}{SNU}.

The analysis accounted for essential system parameters such as the shot noise clearance (see \cref{fig:setup}c), the electronic noise contribution (\SI{0.34}{SNU}), and the quantum efficiency of the receiver ($\eta=0.7$). The \SKR was estimated based on the analytical bound for arbitrary modulation under the trusted noise assumption, incorporating finite-size effects according to:
\begin{equation*}
    \text{SKR} = 2 R_S \frac{n}{N} (1-\text{FER}) (\beta I_{AB} - \chi_{BE} - \Delta),
\end{equation*}
where factor of 2 is due to polarization multiplexing, $R_S$ is the system symbol rate, $I_{AB}$ is the mutual information between Alice and Bob, $ \chi_{BE}$ is the Holevo information between Bob and Eve, and $\Delta$ accounts for the penalty caused by the finite size effects of the privacy amplification~\cite{Milicevic_2018}.
Key operational parameters include a reconciliation efficiency $\beta=96\%$ and a frame error rate (FER) of 50\%, using an expanded protograph-based low-density parity check code ($R~=~1/5$) punctured to the appropriate rate with an error correction blocklength of $10^5$ \cite{gumucs2021low}.
Given that 50\% of the transmitted symbols are pilots and 50\% of the remaining data symbols are utilised for parameter estimation, the fraction of symbols available for key generation is $n/N=0.25$.

\Cref{fig:results}b displays the estimated \SKR as a function of transmission distance.
Theoretical \SKR curves, calculated using the measured system parameters, are shown for total block lengths ($N$) of $2\cdot10^6$ and $2\cdot10^7$, alongside the asymptotic limit ($N\rightarrow\infty$).
The achieved experimental \SKR is highlighted by the red star, demonstrating a key rate of \SI{1.9}{Mbit/s} over \SI{25.2}{km} of standard fibre for a practical block length of $N = 2\cdot10^7$.

\section{Conclusions}

We demonstrated the first \CVQKD system built on top of a custom monolithically integrated dual-polarisation silicon photonic transceiver, showcasing the potential of photonic-electronic integration to enable practical and compact quantum-secure communications.
By tightly co-packaging the silicon \PIC with \TIAs, we achieved a secret key rate of \SI{1.9}{Mbit/s} over \SI{25}{km} of standard single-mode fibre under finite-size analysis.
Further enhancement of the key rates is anticipated through dedicated optimisation of the receiver electronics to reduce excess noise and improve shot noise clearance.
In addition, the next development milestones include investigating other PIC material and packaging systems for laser source integration and deploying the system between two distinct transceivers, which will require mainly the addition of a frequency recovery \DSP stage.
This work represents a key step towards realising low-complexity \CVQKD systems within more compact form factors.

\clearpage
\section{Acknowledgements}
The Dutch Ministry of Economic Affairs (EZ) supported this work under the Quantum Delta NL and the PhotonDelta National Growth Funds Programme. We also acknowledge the support of CUbIQ Technologies via the EIC Transition Grant Project PAQAAL under grant number 101213884.

\printbibliography[]

\vspace{-4mm}

\end{document}